\documentclass[a4paper,12pt]{iopart}

\usepackage[utf8]{inputenc}
\usepackage{graphicx}        
\usepackage{color}

\def \pb {p^{(0)}}
\def \eb {\rho^{(0)}}
\def \Eb {E}
\def \Pb {P}

\def\mo{m_{0}}
\def \ro {a}
\def \Po {P_0^*}
\def \po {p_0^*}
\def \poN{\tilde{p}_0^*}
\def \moN{\tilde{m}_0}

\def \Uo {U^{(0)}}
\def \Us {U^{(2)}}
\def \Uso {U^{(2)}_0}
\def \Mf {M^{(0)}}
\def \Mst {M^{(2)}_T}
\def \Ms {M^{(2)}}

\def \defor {\zeta}

\def \chem {\gamma}

\def \pertp {\varepsilon}

\begin{document}
\title{On the mass of rotating stars in Newtonian gravity and GR}

\author{Borja Reina and Ra\"ul Vera}
\address{Dept. of Theoretical Physics and History of Science,\\ University of the Basque Country UPV/EHU,\\
644 PK, Bilbao 48080, Basque Country, Spain}
\date{}

\begin{abstract}
We show how the correction to the calculation of the mass
in the original relativistic model of a rotating star  by Hartle \cite{Hartle1967}, found recently in \cite{ReinaVeraCiM},
appears in the Newtonian limit, and that the correcting term is
indeed present, albeit hidden, in the original Newtonian approach by
Chandrasekhar \cite{Chandra_Poly_Newton_1933}.
\end{abstract}

\section{Introduction}

The original treatment aimed at the study of (rigidly) rotating stars in a perturbative
scheme is due to Chandrasekhar for Newtonian gravity, back in 1933 \cite{Chandra_Poly_Newton_1933}.
It was only in 1967 that Hartle put forward the model within the realm of
General Relativity \cite{Hartle1967}.
Although the study in \cite{Hartle1967} covers any barotropic equation
of state, 
the work in \cite{Chandra_Poly_Newton_1933} focuses, from some point onwards, only on polytropic equations of state,
i.e. of the form $p=K \rho^{1+\frac{1}{n}}$ for some constants $K$ and $n$, where
$p$ and $\rho$ denote the pressure and the mass density of the star.
The relationship between the Newtonian and the GR approaches was presented in \cite{Hartle1967},
and the GR procedure was found to be consistent with the Newtonian case by taking care of the
suitable limit.

However, it has been found recently \cite{ReinaVeraCiM} that the computation of the total mass of the rotating
configuration as a function of 
the central density shown in \cite{Hartle1967} has to be amended by
a term proportional to the value of the background energy density at the surface of the star.
That value is zero for certain equations of state (including polytropic EOS), but it does not
vanish necessarily (for instance in models of strange quark stars \cite{ColpiMiller}). 
At the time of writing  \cite{ReinaVeraCiM} we did not pursue the study of the Newtonian limit,
since no such term seemed to appear in the literature concerning Newtonian stars.
However, as we show next, that term contributes to the Newtonian limit, and
appears indeed, although implicitly, in the original work
by Chandrasekhar \cite{Chandra_Poly_Newton_1933}.
appearance of that term had been somehow forgotten, even in the review of the Newtonian approach in \cite{Hartle1967}.

\section{The Newtonian star}
We first concentrate on the computation of the mass as stated in \cite{Chandra_Poly_Newton_1933},
expand that for a general case, and show how the expressions in \cite{Chandra_Poly_Newton_1933}
for polytropic equations of state follow indeed.
In page 396 of \cite{Chandra_Poly_Newton_1933} the mass is claimed to be given by
\begin{equation}
  M=2\pi\int\int\rho r^2dr d\mu,
\label{eq:def_mass}
\end{equation}
where $\mu:= \cos\theta$, and $\{r,\theta,\phi\}$ are spherical coordinates, so that $r$ and $\theta$ are the radial coordinate
and azimutal angle on the sphere respectively.

In agreement with the next equation in \cite{Chandra_Poly_Newton_1933}, as we will show later, (\ref{eq:def_mass})
stands for the integral over
the deformed volume.
Indeed, the shape of the star is described in \cite{Chandra_Poly_Newton_1933}
to be the sphere of the background configuration plus a deformation 
at first order in a perturbation parameter $v$, which corresponds to a second order in the
angular velocity  $\omega^2$ over the value of the central density, see (10) in \cite{Chandra_Poly_Newton_1933}.
The value of the central density is denoted by $\lambda$ in \cite{Chandra_Poly_Newton_1933},
but we will use $\rho_c$ here.

Let us first review just the necessary of the Newtonian treatment in order
to obtain the expression of the total mass of the rotating star, suitable to be computed by solving
the relevant problems at different orders in the perturbation.
We follow essentially the description of the Newtonian approach as made in \cite{Hartle1967},
and will compare with that in \cite{Chandra_Poly_Newton_1933} when necessary.

The non-rotating (spherically symmetric) configuration is described by the mass density $\eb(r)$,
pressure $\pb(r)$ and Newtonian potential $\Uo(r)$. The radial variable $r$
runs from $0$ to $\ro$ in the interior and $r>a$ corresponds to the vacuum exterior.
The three equations of structure that govern the configuration are a barotropic equation of state $\pb=\pb(\eb)$, the hydrostatic equilibrium first integral and the Poisson equation, i.e. 
\begin{equation}
\chem = \int_0^{r} \frac{1}{\eb(s)} \frac{d\pb (s)}{ ds}ds + \Uo (r), \quad \nabla^2 \Uo (r) = 4\pi G \eb (r), \label{eq:structure_bck_newt}
\end{equation}
where the constant $\chem$ is identified as the chemical potential.
We use $\nabla^2$ for the flat Laplacian in spherical coordinates $\{r,\theta,\phi\}$. 
In spherically symmetric configurations the mass function is given by
\begin{equation}
\Mf (r)=4\pi\int_0^r   \eb (s)s^2ds.
\label{eq:mass_bck_newt}
\end{equation}%
Given that $\Mf(0)=0$ and that regularity at the origin $r=0$ implies $d \Uo/d r(0)=0$, the mass and the potential are related by
\[
\frac{d \Uo(r)}{d r} = \frac{G }{r^2}\Mf(r).
\]
The system of three equations can be integrated in terms of boundary conditions at the origin and thus provide,
e.g. , the total mass of the star, $\Mf_S:=\Mf(\ro)$, in terms of the central density $\rho_c=\eb(0)$.
We denote that function by $\Mf_S(\rho_c)$.

Consider now the (perturbed) rotating configuration. The  mass density of the rotating configuration $\rho(r,\mu)$ is expanded perturbatively to first order in $v$ as
\[
\rho(r,\mu)=\rho^{(0)}(r)+v \rho^{(2)}(r,\mu)+\mathcal{O}(v^2).
\]
A new radial coordinate $R$ is now chosen so that it labels surfaces of constant density in the rotating configuration
by  (see \cite{Hartle1967})
\begin{equation}
\rho(r(R,\mu),\mu)=\eb(R).
\label{eq:def_R}
\end{equation}
The interior of the rotating star is therefore defined by $R\in (0,\ro)$ by construction,
and its surface located at $R=\ro$.  
The change between $R$ and $r$
must thus have the form
\begin{equation}
r(R,\mu)=R + v\defor(R,\mu)+\mathcal{O}(v^2)
\label{eq:change_defor}
\end{equation}
for some function $\defor(R,\mu)$,
which thus describes the deformation of the surface \cite{Chandra_Poly_Newton_1933,Hartle1967}.
Before considering the equations governing the rotating configuration, in the next section,
let us expand the integral (\ref{eq:def_mass}),
which reads explicitly
\begin{equation}
  M=2\pi\int_{-1}^1\int^{\ro+v \defor(\ro,\mu)}_0 \left(\rho^{(0)}(r)+v \rho^{(2)}(r,\mu)\right)r^2dr d\mu+\mathcal{O}(v^2).
\label{eq:mass_bo}
\end{equation}
In \cite{Chandra_Poly_Newton_1933} the radial coordinate $r$
is, instead, conveniently rescaled to a new radial coordinate $\xi$ (eq. (9) in \cite{Chandra_Poly_Newton_1933}).
The shape of the star is then described in \cite{Chandra_Poly_Newton_1933}
to be the sphere of the background configuration $\xi_1$ plus a deformation denoted by $d\xi_1$
at first order in $v$.
For a polytropic equation of state $P=K \rho^{1+\frac{1}{n}}$, $\xi_1$ is the first zero of Emden's function
with index $n$, denoted by $\theta(\xi)$ in \cite{Chandra_Poly_Newton_1933}, and the deformation corresponds to $d\xi(\xi_1,\mu)$
for some function $d\xi(\xi,\mu)$ that can be extracted from the terms in the $v$ factors in equations (36) and (38)
in \cite{Chandra_Poly_Newton_1933}.
Obviously $v\defor(R,\mu)$ scales to $d\xi(\xi,\mu)$ as $r$ scales to $\xi$ 
(9) in \cite{Chandra_Poly_Newton_1933}.
Note that $d\xi$ contains $v$.
After using (7), (9), (36) in \cite{Chandra_Poly_Newton_1933} for the polytropic equation of state, (\ref{eq:mass_bo})
can be shown (see below) to translate, up to order $v$, to
\begin{equation}
M=4\pi\left[\frac{(n+1)K}{4\pi G} \lambda^{\frac{1}{n}-1}\right]^{3/2}\lambda\int^{\xi_1+d\xi_1}_0(\theta^n+v n\theta^{n-1}\psi_0)\xi^2d\xi,
\label{eq:mass_chandra_poly}
\end{equation}
as it stands in page 396 in \cite{Chandra_Poly_Newton_1933}, where $d\xi_1$ denotes the $l=0$ part of $d\xi(\xi_1,\mu)$,
which equals $d\xi_1=-v\psi_0(\xi_1)/\theta'(\xi_1)$\footnote{Since $\theta'(\xi_1)<0$ \cite{Chandra_Poly_Newton_1933},
$-\theta'(\xi_1)$ always appears as $|\theta'(\xi_1)|$ in \cite{Chandra_Poly_Newton_1933}.}, see (38) in \cite{Chandra_Poly_Newton_1933},
where $\psi_0(\xi)$ is an auxiliary function that satisfies (37${}_1$) in \cite{Chandra_Poly_Newton_1933}.
$d\xi_1$ is the expansion of the star, as noted in \cite{Chandra_Poly_Newton_1933}.
Only the $l=0$ sector contributes to the integral.

In order to obtain (\ref{eq:mass_chandra_poly}) and go further let us develop (\ref{eq:mass_bo}).
Since the Jacobian of the change (\ref{eq:change_defor}) is  $1+v \partial \defor/\partial R$, 
the integral (\ref{eq:mass_bo}) expands as
\fl
\begin{eqnarray}
\fl
M=2\pi\int_{-1}^1\int^{\ro}_0 \rho(r(R,\mu),\mu)(R^2+2v\defor R)(1+v\frac{\partial \defor}{\partial R}) dRd\mu+\mathcal{O}(v^2)\nonumber\\
\fl
=2\pi\int_{-1}^1\int^{\ro}_0 \eb(R)
(R^2+2v\defor R)(1+v\frac{\partial \defor}{\partial R}) dRd\mu+\mathcal{O}(v^2)\nonumber\\
\fl
=2\pi\int_{-1}^1\int^{\ro}_0 \left[\eb(R)R^2
+v\left(2\rho^{(0)}(R)\defor R+R^2\eb(R)\frac{\partial \defor}{\partial R}\right)\right]dRd\mu+\mathcal{O}(v^2) \nonumber\\
\fl
=2\pi\int_{-1}^1\int^{\ro}_0 \left[\eb(R)R^2+v
\left( -R^2\defor \frac{d \eb}{d R}+\frac{\partial}{\partial R}\left(R^2\eb(R)\defor\right)\right)\right]dRd\mu+\mathcal{O}(v^2)\nonumber\\
\fl
=4\pi\int^{\ro}_0 \eb(R)R^2dR-2\pi v \int^1_{-1}\int^\ro_0 R^2\defor \frac{d\eb }{d R}dRd\mu
+2\pi v\int^1_{-1}\ro^2\eb(\ro)\defor(\ro,\mu)d\mu \label{eq:mass_final1}\\
+\mathcal{O}(v^2),\nonumber
\end{eqnarray}
where the relation (\ref{eq:def_R}) that defines $R$ has been used in the second equality.
The first term in the final expression (\ref{eq:mass_final1})
corresponds to $\Mf(\ro)$ by (\ref{eq:mass_bck_newt}). The second term is more easily
recognised by developing the relation (\ref{eq:def_R}) as follows.
Given that for any $f(r,\mu)$ we have
\begin{equation}
f(r(R,\mu),\mu)=f(R,\mu)+v f'(R,\mu) \defor +\mathcal{O}(v^2),
\label{eq:f_in_defor}
\end{equation}
where the prime denotes differentiation with respect to the first (or only) argument, the relation
(\ref{eq:def_R}) provides, in particular,
\begin{equation}
  \rho^{(2)}(R,\mu)=-\defor(R,\mu) \frac{d\eb(R) }{d R}.
\label{eq:rho2}
\end{equation}
The expression (\ref{eq:mass_final1}) can thus be also written as
\begin{eqnarray}
M&=&4\pi\int^{\ro}_0 \rho^{(0)}(R)R^2dR+2\pi v \int^1_{-1}\int^\ro_0 \rho^{(2)}(R,\mu)R^2dRd\mu\nonumber\\
 &&+2\pi v\int^1_{-1}\ro^2\rho^{(0)}(\ro)\defor(\ro,\mu)d\mu+\mathcal{O}(v^2).\label{eq:mass_final1b}
\end{eqnarray}

From now onwards let us denote by a $f_0$ (subindex ${}_0$)
the part of any function $f$
parallel 
to the Legendre polynomial $P_0(\mu)(=1)$. 
In other words, $f_0(\cdot):=\frac{1}{2}\int f(\cdot,\mu) P_0(\mu)d\mu$.
We will also refer to $f_0$ as the $l=0$ sector of $f$. 
The mass (\ref{eq:mass_final1b}) thus reads
\begin{equation}
\fl
M=4\pi\int^{\ro}_0 \rho^{(0)}(s)s^2ds+4\pi v \int^\ro_0 \rho^{(2)}_0(s)s^2ds
+4\pi v\ro^2\rho^{(0)}(\ro)\defor_0(\ro)+\mathcal{O}(v^2).
 \label{eq:mass_final2}
\end{equation}
The fact that only the $l=0$ sector contributes to the integral is now explicit.

For polytropic equations of state, after using (7), (9), (36) and (38)  in \cite{Chandra_Poly_Newton_1933},
(\ref{eq:mass_final2}) directly translates, up to order $v$, to
\begin{equation}
\fl
M=4\pi\left[\frac{(n+1)K}{4\pi G} \lambda^{\frac{1}{n}-\frac{1}{3}}\right]^{3/2}
\left\{\int^{\xi_1}_0\theta^n\xi^2d\xi+
v \int_0^{\xi_1}n\theta^{n-1}v\psi_0\xi^2d\xi- v\xi_1^2 \theta^n(\xi_1)\frac{\psi_0(\xi_1)}{\theta'(\xi_1)} \right\},
\label{eq:mass_chandra_2}
\end{equation}
which is not difficult to show 
to be equivalent to  (\ref{eq:mass_chandra_poly}) irrespective of the equation that the function $\theta(\xi)$
satisfies.

The crucial point here is, let us recall,  that the function $\theta(r)$ is Emden's function, for which $\theta(\xi_1)=0$
by construction, which is equivalent to $\rho^{(0)}(\ro)=0$.
The above expression (\ref{eq:mass_chandra_2})
for the total mass is obviously presented in \cite{Chandra_Poly_Newton_1933} without the last term, which vanishes
(see above (40) in \cite{Chandra_Poly_Newton_1933}).
However, in general, the mass density $\rho^{(0)}(R)$ of the background spherical
configuration does not have to vanish necessarily at the boundary $R=\ro$. The expression of the total mass
in \cite{Chandra_Poly_Newton_1933}, made explicit for a class of equations of state for which the mass density vanishes at the
surface of the star, seems to have misled many authors to forget the third term in (\ref{eq:mass_final2}).
Even the author himself forgot, many years later, to include that term when exploring homogeneous (constant $\rho$) stars
in GR \cite{Chandra_Miller1974}. The correction to the calculation of the mass of homogeneous stars
can now be found in \cite{Borja_constant_rho}.

The third term in (\ref{eq:mass_final2}), proportional to $\rho^{(0)}(\ro)$,
corresponds, precisely, to the Newtonian limit of the term found in \cite{ReinaVeraCiM}
that amends the ``change in mass'' computed in \cite{Hartle1967}. That is shown in the following
section, where we very briefly review the equations for the perturbed configuration needed in both Newtonian gravity and GR.

\section{The mass in Newtonian gravity and GR}
\subsection{Newtonian gravity}
We only need to consider the $l=0$ sector of the perturbation.
As in the background configuration, apart from the given barotropic equation of state,
the perturbation at first order in $v$
is governed by a hydrostatic equilibrium first integral and a Poisson equation
\begin{eqnarray}
&&\Uso (R) + \defor_0 (R) \frac{d \Uo (R)}{dR} - \frac{2 \pi G \rho_c R^2}{3} = 0, \label{eq:hydro_so_newt}\\
&&\frac{1}{R^2}\frac{d}{d R}\left( R^2 \frac{d \Uso (R)}{d R}\right)
=  -4 \pi G \defor_0 (R) \frac{ d \eb(R)}{dR}, \label{eq:poisson_so_newt}
\end{eqnarray}
where the Poisson equation for the nonrotating  potential has been used in the second equality.
Note that from (\ref{eq:rho2}) we have $\rho^{(2)}_0(R)=-\defor_0(R) d \eb(R)/dR$,
so that the right hand side of (\ref{eq:poisson_so_newt}) can be also expressed as $4\pi G\rho^{(2)}_0(R)$.
It is important to note that the domains of definition of these equations are given by $R\in(0,\ro)$
for the interior and $R>\ro$ for vacuum, and suitable boundary conditions
(including a regular origin and asymptotic flatness) are imposed accordingly.

It is convenient to change the functions $\{\Uso, \defor_0\}$
that describe the configuration to a new set $\{\Ms, \po\}$, suitable to be compared with the relativistic model,
defined as follows, 
\begin{eqnarray}
&&\Ms(R):=4\pi  \int^R_0 \rho^{(2)}_0(s)s^2ds=-4\pi  \int^R_0 \defor_0(s) \frac{d\eb(s) }{d s}s^2ds,
\label{defi:m_pert_newt_bad}\\
&&\po (R) := \frac{G M^{(0)}(R)}{R^2}\defor_0(R). 
\label{defi:po_xi}
\end{eqnarray}
The definition (\ref{defi:po_xi}) can be expressed in terms of the pressure and
density of the background configuration by
differentiating the hydrostatic equilibrium first integral for the
static configuration (first equation in
(\ref{eq:structure_bck_newt})), which provides
\[
\frac{d\Uo(R)}{dR} + \frac{1}{\eb(R)}\frac{d\pb(R)}{dR} = 0,
\]
so that 
\begin{equation}
\defor_0 (R) = -\eb(R)\left(\frac{d\pb(R)}{dR}\right)^{-1} \po (R). \label{xi_to_po}
\end{equation}
On the other hand, the second order Poisson equation (\ref{eq:poisson_so_newt}) can be expressed
in terms of the pressure perturbation factor by using (\ref{xi_to_po}) to get
\begin{equation}
\frac{1}{R^2}\frac{d}{d R}\left( R^2 \frac{d \Uso (R)}{d R}\right)= 4 \pi G \frac{d \eb}{d \pb}\eb  \po(R). \label{eq:poisson_so_newt_2}
\end{equation}
We can also rewrite the expression for $\Ms$, (\ref{defi:m_pert_newt_bad}),
using (\ref{xi_to_po}) and (\ref{defi:po_xi}), which in differential form reads (see (15) in  \cite{Hartle1967})
\begin{equation}
\frac{d \Ms(R) }{dR} = 4 \pi  R^2  \frac{d \eb}{d\pb}\eb \po(R).   \label{eq:m_pert_newt}
\end{equation}
The equation for $\po$ is obtained as follows. Combine (\ref{eq:poisson_so_newt_2}) with (\ref{eq:m_pert_newt})
to get rid of $\po$ and integrate once 
taking into account that $\Ms (0) = 0$, by construction, and $d\Uso/dR|_{R=0} = 0$
for a regular origin. We thus obtain
\begin{equation}
 \frac{d\Uso(R)}{dR}= \frac{G}{R^2}\Ms(R), \label{first_derivative_Uso}
\end{equation}
in anlogy with the background configuration.
Finally, take the derivative of the  hydrostatic equilibrium first integral (\ref{eq:hydro_so_newt})
\begin{equation}
\frac{d \Uso(R) }{dR} + \frac{d}{dR}\left( \defor_0 (R)\frac{d\Uo (R)}{dR}\right) - \frac{4 \pi G \rho_c}{3}R = 0,
\end{equation}
and use (\ref{first_derivative_Uso}) and (\ref{defi:po_xi}) to obtain (see (15) in \cite{Hartle1967})
\begin{equation}
\frac{d\po(R)}{dR} = - \frac{G}{R^2}\Ms(R)  + \frac{4\pi G \rho_c}{3} R. \label{eq:po_newt}
\end{equation}
The system of equations for the functions $\{\Ms, \po\}$ is formed by (\ref{eq:m_pert_newt}) and (\ref{eq:po_newt})
on the domain $R\in(0,\ro)$.
As in the background configuration system, this problem allows us to integrate $\{\Ms, \po\}$ given
boundary conditions at the origin. In particular one can compute $\Ms_S:=\Ms(\ro)$ as a function
of the (total) central density $\hat\rho_c$, and thus construct a function $\Ms_S(\hat\rho_c)$.
In order to add this function to the contribution from the background configuration  $\Mf_S(\rho_c)$
it is, of course, necessary to choose $\hat\rho_c=\rho_c$, so that $\rho_c$ becomes
a parameter of the whole perturbed configuration. That implies choosing
$\po(0) = 0$.

The total mass of the rotating configuration (\ref{eq:mass_final2}), taking into account
 (\ref{eq:mass_bck_newt}), (\ref{defi:m_pert_newt_bad}) and (\ref{defi:po_xi}), can be expressed as
\begin{equation}
 M=\Mf(\ro) +v\Ms(\ro)+4\pi v  \frac{\ro^4}{G\Mf(\ro)} \rho^{(0)}(\ro)\po(\ro)+\mathcal{O}(v^2).
\label{eq:newtonian_total_mass}
\end{equation}
Note, again, that this sum makes sense once the functions involved are computed given common boundary
data, in terms of a common set of parameters, as for instance $\rho_c$.
Nevertheless, the choice of parameter used to compute those functions is irrelevant for our purposes.
The contribution of the perturbation to the total mass in Newtonian gravity is given by
\begin{equation}
  \label{eq:Newton_CiM}
  \Mst=\Ms(\ro)+4\pi \frac{\ro^4}{G\Mf(\ro)} \rho^{(0)}(\ro)\po(\ro).
\end{equation}
The second term in the above expression is missing in the first equality of equation (18) in \cite{Hartle1967}.

\subsection{General Relativity}
The metric up to second order in some parameter $\pertp$,
following the notation in \cite{Hartle1967} (see also \cite{ReinaVeraCiM}),  reads
\begin{eqnarray}
g_\pertp &=& -e^{\nu(r)} (1+2 \pertp^2 h(r, \theta))dt^2 + e^{\lambda(r)}\left(1+2\pertp^2 \frac{m(r,\theta)}{r-2M}\right) dr^2 \nonumber\\
&&+ r^2 (1+2 \pertp^2 k(r,\theta))\left( d \theta^2 + \sin^2 \theta (d \varphi - \pertp \omega(r)dt)^2\right).\nonumber
\end{eqnarray}
Given that we follow \cite{Hartle1967} and \cite{ReinaVeraCiM} in this section, we also use geometrized units for convenience, so that $G=c=1$ unless otherwise stated. We can fix the (dimensionless) perturbation parameter $\pertp$ in analogy with the formalism developed in  \cite{Chandra_Poly_Newton_1933} for the Newtonian model. To this aim we set $\pertp^2 = v =  \omega^2/2\pi E(0)$, where $E(0)$ is the energy density of the background configuration at the origin, and $\omega$ is the constant angular velocity of the fluid, as in the Newtonian treatment. Therefore, the quantity that drives the perturbations in \cite{Hartle1967} is expressed here by $\Omega^H = \sqrt{2 \pi E(0) v}$, whereas the constant $\Omega$ used in \cite{ReinaVeraCiM}  is identified with $\Omega = \sqrt{2\pi E(0)}$ in this convention.\footnote{This value is related to the usual choice in numerical works (see e.g. \cite{ColpiMiller}),
in which an estimate of the mass shedding frequency is chosen, say, $\Omega^*=\sqrt{M/a^3}$.
For a constant density star it is easy to check that $\sqrt{2 \pi E(0)} = \sqrt{3M/2a^3}=\sqrt{3/2}\,\Omega^*$.}

We shall keep the perturbation parameter $\pertp$ and the constant $\Omega$ in this section in order to ease the comparison with \cite{ReinaVeraCiM}, although the identifications will be made explicit when the Newtonian limit is taken.

As in the Newtonian case, we only need focusing on the $l=0$ sector of the solution for our purposes. The coordinate
$r$ is fixed by choosing $k_0(r)=0$ \cite{Hartle1967} (see also \cite{ReinaVeraCiM} for a discussion on
the choice of gauges).
The asymptotically flat vacuum solution is given by \cite{Hartle1967}
\begin{eqnarray}
e^{\nu_{vac}(r)} = 1-\frac{2M}{r}=e^{-\lambda_{vac}(r)},\quad \omega^{vac}(r) = \frac{2J}{r^3}, \nonumber\\
h^{vac}_0(r) = -\frac{ \delta M}{r-2M} + \frac{J^2}{r^3(r-2M)}, \quad \mo^{vac} (r) = \delta M - \frac{J^2 }{r^3},\label{eq:m0_vacuum}
\end{eqnarray}
where $M$, $J$ and $\delta M$ are constants. In the analysis of the background and first order
configurations,
$M$ and $J$ are identified as the background mass and the angular momentum, respectively.
The equations governing the background and first order configurations are used
to compute $M$ and $J$ given suitable data at the origin. 
We refer to \cite{Hartle1967} for a full account (see also \cite{Bradley_etal2007,ReinaVeraCiM}).
The constant $\delta M$, still to be determined, is identified with the ``change in mass'' due to the second order perturbation,
or simply the contribution to the mass at second order, since  the angular independent part of $g_{tt}$ is given by
\begin{equation}
\left. g_{tt} \right|_{r \to \infty} = 1 - \frac{2}{r} (M + \pertp^2 \delta M) + \mathcal{O}\left(\frac{1}{r^2} \right).
\end{equation}
The  $l=0$ sector of the (second order) perturbation interior configuration is completely determined
by the pair of functions $\{\mo(r), \Po(r)\}$, with
\begin{equation}
\Po := \frac{P^{(2)}_0}{2 (\Eb + \Pb)},
\end{equation}
where $\Eb$ and $\Pb$ are the energy density and pressure of the static background interior, respetively, and $P^{(2)}(r,\theta)$
the perturbation to the pressure (see \cite{ReinaVeraCiM} for this alternative definition of the
same function $\po$ in \cite{Hartle1967}).
The system of equations that $\{\mo, \Po \}$ satisfy are to be fulfilled in the domain $r\in(0,\ro)$,
with suitable boundary conditions, and read \cite{Hartle1967,ReinaVeraCiM}
\begin{eqnarray}
\fl\frac{d\mo}{dr} = 4 \pi r^2 (E+P)\frac{dE}{dP}\Po + \frac{1}{12} j^2 r^4 \left(\frac{d\tilde{\omega}}{dr} \right)^2 - \frac{2}{3} r^3 j \frac{dj}{dr} \tilde{\omega}^2, \label{eq:efe_mo}\\
\fl\frac{d\Po}{dr} = -4 \pi \frac{(E+P)r^2}{r-2M}\Po - \frac{r^2\mo}{(r-2M)^2}   + \frac{1}{12} \frac{r^4j^2}{r-2M}  \left( \frac{d\tilde{\omega}}{dr}\right)^2 +\frac{1}{3} \frac{d}{dr} \left(\frac{r^3j^2 \tilde{\omega}^2}{r-2M} \right), \label{eq:efe_po}
\end{eqnarray}
where $\tilde\omega(r):=\omega(r)-\Omega$ and $j(r):= \exp[-(\nu+\lambda)/2]$.

The value of $\delta M $ is determined in terms of interior quantities
using the matching conditions for the exterior and interior problems to second order provided in
\cite{ReinaVeraCiM}.
In particular, a function $\mo(s)$ for $s\in(0,\infty)$ constructed by joining $\mo(s)$ and $\mo^{vac}(s)$ across
$s=\ro$ \emph{is not continuous in general}, since it presents a jump proportional to $E(\ro)$.
The result is  \cite{ReinaVeraCiM}
\begin{equation}
\delta M =  \mo (\ro) + \frac{J^2}{\ro^3} + 4 \pi \ro^3 \frac{\ro-2M}{M} \Eb(\ro) \Po(\ro). \label{deltaM}
\end{equation}
The third term  accounts precisely for the jump in $\mo$.
Let us stress here that this term does not appear in the analysis in \cite{Hartle1967}
because the function $\mo(s)$ is assumed, \emph{a priori}, to be continuous across $s=a$. The relationship
between the continuity of $\mo$ and the vanishing of $E(\ro)$, and therefore the validity of the assumption
for certain equations of state (in particular the polytropes) is proven and discussed in \cite{ReinaVeraCiM},
using the fully consistent theory of perturbed matching to second order in \cite{Mars2005}.

As in the Newtonian case, the background quantities
$ E(\ro)$, $M$ and $J$, and the perturbation ones, $\mo(\ro)$ $ \Po(\ro)$ are to be computed by solving the corresponding
system of equations given the (common) relevant data at the origin.
In \cite{Hartle1967} the parameter chosen is the central desnsity
$\rho_c$, but, as mentioned above, that choice is not relevant for this discussion.

\subsection{Newtonian limit}
Our purpose now is to obtain
the Newtonian limit of $\delta M$ in (\ref{deltaM}) and compare it with the contribution to the mass of the perturbation
in the Newtonian approach, $\Mst$, given by (\ref{eq:Newton_CiM}).
First, though, it is convenient to find the Newtonian limit for the system (\ref{eq:efe_mo}) and (\ref{eq:efe_po})
in order to relate $\{\mo, \Po \}$ with the pair $\{\Ms, \po \}$ from the Newtonian approach. This is achieved by
performing an expansion in powers of $1/c$ as (see \cite{Hartle1967})
\begin{eqnarray}
\fl
M = \frac{G}{c^2}\Mf_S + \mathcal{O}\left( \frac{1}{c^4}\right),\quad E(r) =  \frac{G }{c^2}\eb(r) + \mathcal{O}\left( \frac{1}{c^4}\right),
\quad P(r)=\frac{G}{c^4} \pb(r) + \mathcal{O}\left(\frac{1}{c^6} \right),\nonumber\\
\fl
\tilde{\omega}(r) = -\frac{\sqrt{2\pi G \rho_c}}{c} + \mathcal{O}\left( \frac{1}{c^3}\right),\nonumber\\
\fl\Po(r) = \frac{1}{c^2}\poN(r) + \mathcal{O}\left(\frac{1}{c^4}\right), \quad \mo(r) = \frac{G}{c^2}\moN(r) + \mathcal{O}\left( \frac{1}{c^4}\right), \nonumber
\end{eqnarray}
for some functions $\moN$ and $\poN$, 
where $\eb$ (and $\rho_c$), $\pb$ and $\Mf$ correspond to the functions describing the Newtonian background configuration.
Note that, concerning the first order (in $\pertp$), the function $\tilde\omega(r)$ is constant at lowest order in $1/c$
\cite{Hartle1967}. 
Given the system (\ref{eq:efe_mo}) and (\ref{eq:efe_po}), the pair $\{\moN,\poN\}$ thus satisfies (see (102)-(103) in \cite{Hartle1967})
\begin{eqnarray}
\frac{d \moN}{dr} &=& 4\pi r^2 \frac{d\eb}{d\pb} \rho \poN, \\
\frac{d \poN}{dr} &=& - \frac{G\moN}{r^2} + \frac{4\pi G \rho_c}{3} r.
\end{eqnarray}
Compare this system of equations with (\ref{eq:m_pert_newt}) and (\ref{eq:po_newt}).
The functions arising from the newtonian limit $\{\moN, \poN\}$ and the functions in the
perturbed Newtonian model $\{\Ms, \po \}$ satisfy the same equations 
\emph{in the same domain} $r,R \in (0, \ro)$. Therefore,
the pair  $\{\moN,\poN\}$ is equivalent to $\{\Ms, \po \}$ for $r,R<\ro$.
We can now substitute $\{\moN,\poN\}$ by $\{\Ms, \po \}$ in the following.

The Newtonian limit for (\ref{deltaM}) is obtained following procedure above together with
\begin{equation}
\delta M = \frac{G}{c^2} \widetilde{\delta M} + \mathcal{O}\left(\frac{1}{c^4} \right),
\end{equation}
for some $ \widetilde{\delta M}$, from where (\ref{deltaM}) becomes
\begin{equation}
\widetilde{\delta M} = \Ms(\ro) + 4 \pi  \frac{\ro^4 }{G\Mf_S}\eb(\ro)\po (\ro) + \mathcal{O}\left(\frac{1}{c^4} \right) . \label{deltaM_newtonianlimit}
\end{equation}
Comparing this expression with (\ref{eq:Newton_CiM}) we finally find
\[
\widetilde{\delta M}=\Mst,
\]
that is, the Newtonian limit of the contribution (to second order) of the perturbation to the mass
in GR, found in \cite{ReinaVeraCiM}, is non zero, and agrees with the same quantity computed in Newtonian gravity.

\section{The Newtonian matching conditions}
As a final remark, let us comment
on the boundary conditions at the surface of the star,
the matching between the interior and exterior problems at each order,
involved in the Newtonian approach.
Some objections to the Newtonian matching problem stated in \cite{Chandra_Poly_Newton_1933}
were raised in \cite{Jardetzky_58}. Those were finally solved by Chandrasekhar and Lebovitz in \cite{Chandra_Lebovitz_III_62}
by properly formulating the matching and producing the same results.
However, \cite{Chandra_Lebovitz_III_62} concerns, again, only polytropic equations of state, and
the matching conditions are obtained only for that case, which in particular satisfies $\rho(\ro)=0$.
Let us, for completeness, deduce the matching conditions for the perturbed Newtonian potential
in the general case,
which, as expected, turns out to be compatible with the obtaining of the perturbed mass
(\ref{eq:newtonian_total_mass}).

Consider any function $f$ that depends on $v$ on two arguments by $f(\ro+v\defor(\ro,\mu),\mu,v)$.
Let us use $\partial_v$ to denote a derivative with respect to the third argument, and
a prime $'$ with respect to the first, and define
$f^{(0)}(a,\mu):=f(\ro+v\defor(\ro,\mu),\mu,v)|_{v=0}=f(\ro,\mu,0)$, and 
$f^{(2)}(a,\mu):=\partial_vf(\ro+v\defor(\ro,\mu),\mu,v)|_{v=0}$.
Assume now that $f$ satisfies the equation
\[
f(\ro+v\defor(\ro,\mu),\mu,v)=0.
\]
Evaluating the equation at $v=0$ we obtain
\begin{equation}
f^{(0)}(a,\mu)=0,
\label{eq:eq_f}
\end{equation}
while differentiating with respect
to $v$, and then evaluating at $v=0$ we get
\begin{equation}
f^{(2)}(a,\mu)+f^{(0)}{}'(\ro) \defor(a,\mu)=0.
\label{eq:eq_df}
\end{equation}

Consider now two families of interior and exterior problems, defined by some parameter $v$,
with their respective Newtonian potentials $U_{int}(r,\mu,v)$ and $U_{ext}(r,\mu,v)$.
The matching of the problems at each $v$ accounts for the equality of the potentials
and their radial derivatives at the common boundary, the matching surface. Let us define
the matching surface by $r(a,\mu)=a+v\defor(a,\mu)$. The conditions are, therefore,
 $U_{int}(a+v\defor(a,\mu),\mu,v)=U_{ext}(a+v\defor(a,\mu),\mu,v)$
and  $U_{int}'(a+v\defor(a,\mu),\mu,v)=U_{ext}'(a+v\defor(a,\mu),\mu,v)$.
Let us finally use the notation $[g]:=g_{int}-g_{ext}$, so that the matching conditions read
$[U](a+v\defor(a,\mu),\mu,v)=0$ and $[U'](a+v\defor(a,\mu),\mu,v)=0$.
These two functions satisfy the requirements for $f$ above, so 
it is now just a matter of applying equations
(\ref{eq:eq_f}) and (\ref{eq:eq_df}) for both
$[U]$ and $[U']$. The four equations thus obtained read
\begin{eqnarray}
  \fl[\Uo](\ro)=0, \qquad [\Uo{}'](\ro)=0,\nonumber\\
  \fl[\Us](\ro,\mu)=-[\Uo{}'](a)\defor(\ro,\mu)=0, \qquad [\Us{}'](\ro,\mu)=-[\Uo{}''](a)\defor(\ro,\mu),
\label{eq:Newt_matching}
\end{eqnarray}
where we have used that the background potentials $\Uo$'s do not depend on $\mu$. 
The last equation in (\ref{eq:Newt_matching}) yields
\begin{equation}
  \label{eq:Newt_jump}
  [\Us{}'](\ro,\mu)=-4\pi G \eb(a)\defor(\ro,\mu).
\end{equation}

As expected, although the background potential, its first derivative
and the potential at first order are `continuous',
the radial derivative of the Newtonial potential at first order in $v$ suffers a jump,
proportional to $\eb(a)$.
The perturbed mass can now be computed from the Newtonian potential,
and it is straightforward to show that this jump generates
the term proportional to $\eb(a)$ in (\ref{eq:newtonian_total_mass}).

Finally, it can be shown that the matching condition for $h'$
in \cite{ReinaVeraCiM}, which suffers a jump proportional to the jump of $m$,
agrees with (\ref{eq:Newt_matching}) after taking the Newtonian limit.

\ack
We acknowledge financial support from projects IT592-13 (GIC12/66)
of the Basque Government, FIS2010-15492 from the MICINN, and UPV/EHU under program UFI 11/55.
BR thanks the Basque Government through grant BFI-2011-250.

\section*{References}
\bibliography{references}{}
\bibliographystyle{review_bib}

\end{document}